\documentstyle[epsfig]{aipproc}

\def\MPL #1 #2 #3 {Mod.~Phys.~Lett.~{\bf#1},\  #2 (#3)}
\def\NPB #1 #2 #3 {Nucl.~Phys.~{\bf B#1},\  #2 (#3)}
\def\PLB #1 #2 #3 {Phys.~Lett.~{\bf B#1},\  #2 (#3)}
\def\PR #1 #2 #3 {Phys.~Rep.~{\bf#1},\ #2 (#3)}
\def\PRD #1 #2 #3 {Phys.~Rev.~{\bf D#1},\  #2 (#3)}
\def\PRL #1 #2 #3 {Phys.~Rev.~Lett.~{\bf#1},\  #2 (#3)}
\def\RMP #1 #2 #3 {Rev.~Mod.~Phys.~{\bf#1},\  #2 (#3)}
\def\ZPC #1 #2 #3 {Z.~Phys.~{\bf C#1},\  #2 (#3)}
\def\IJMP #1 #2 #3 {Int.~J.~Mod.~Phys.~{\bf#1},\  #2 (#3)}

\def\vev#1{\langle #1 \rangle}
\def\bit{\begin{itemize}}
\def\eit{\end{itemize}}
\def\pzero{P^0}
\def\mpzero{m_{\pzero}}

\newcommand{\beq}{\begin{equation}}
\newcommand{\eeq}{\end{equation}}
\newcommand{\bea}{\begin{eqnarray}}
\newcommand{\eea}{\end{eqnarray}}
\newcommand{\nn}{\nonumber}

\def\Eq#1{Eq.~(\ref{#1})}

\def\lsim{\mathrel{\raise.3ex\hbox{$<$\kern-.75em\lower1ex\hbox{$\sim$}}}}
\def\gsim{\mathrel{\raise.3ex\hbox{$>$\kern-.75em\lower1ex\hbox{$\sim$}}}}

\def\snu{\wt\nu}

\def\mcnone{m_{\cnone}}

\def\wtil{\widetilde}

\def\to{\rightarrow}

\def\mupmum{\mu^+\mu^-}
\def\cnone{\chi^0}
\def\mcnone{m_{\cnone}}

\def\snt{\tilde\nu_\tau}

\def\msnt{m_{\snt}}
\def\dm{\Delta\msnt}

\def\gamsnt{\Gamma_{\snt}^{\rm tot}}
\def\gev{~{\rm GeV}}
\def\sigsntbar{\overline \sigma_{\snt}}
\def\rp{\not{\hbox{\kern-4pt $R_P$}}}
\def\met{\not{\hbox{\kern-4pt $E_T$}}}
\def\mpt{\not{\hbox{\kern-4pt $p_T$}}}
\def\fbi{ \text{ fb}^{-1}}
\def\anti{\overline}
\def\rts{\sqrt s}
\def\sigrts{\sigma_{\!\rts}}

\def\rslash{/ \hskip-8pt R}
\def\etmiss{/ \hskip-8pt E_T}

\def\what{\widehat}
\def\wtil{\widetilde}

\def\To{\Rightarrow}

\def\etc{{\it etc.}}

\def\etc{{\em etc.}}

\def\calo{{\cal O}}
\def\eg{{\it e.g.}}
\def\etal{{\it et al.}}

\def\mstop{m_{\stop}}

\def\stop{\wt t}

\def\hsm{h_{\rm SM}}
\def\mhsm{m_{\hsm}}
\def\hl{h^0}
\def\hh{H^0}
\def\ha{A^0}
\def\hp{H^+}
\def\hm{H^-}
\def\hpm{H^{\pm}}
\def\mhl{m_{\hl}}
\def\mhh{m_{\hh}}
\def\mha{m_{\ha}}

\def\mhpm{m_{\hpm}}
\def\tanb{\tan\beta}

\def\mt{m_t}

\def\mz{m_Z}
\def\mw{m_W}

\def\wm{W^-}

\def\cnone{\wt\chi^0_1}

\def\snu{\wt\nu}

\def\mcnone{m_{\cnone}}

\def\h{h}

\def\mh{m_{\h}}

\def\wt{\widetilde}

\def\dmm{\Delta^{--}}
\def\mdmm{m_{\dmm}}

\def\gamdmm{\Gamma_{\dmm}^{\rm tot}}

\def\mVV{M_{VV}}

\def\em{e^-}

\def\hpm{H^{\pm}}

\def\call{{\cal L}}
\def\cmsi{{\rm cm}^{-2}{\rm s}^{-1}}
\def\wtil{\widetilde}
\def\tauptaum{\tau^+\tau^-}
\def\mbb{m_{b\anti b}}

\def\ltot{L_{\rm tot}}

\def\lam{\lambda}
\def\lampr{\lambda^{\prime}}
\def\br{B}
\def\tauptaum{\tau^+\tau^-}
\def\mbb{m_{b\anti b}}

\def\shat{{\hat s}}
\def\rtshat{\sqrt{\shat}}
\def\gam{\gamma}
\def\sigrts{\sigma_{\tiny\rts}^{}}

\def\etal{{\it et al.}}
\def\etc{{\it etc.}}
\def\sighbar{\overline \sigma_{\h}}

\def\sighsmbar{\overline\sigma_{\hsm}}

\def\anti{\overline}
\def\epem{e^+e^-}
\def\zstar{Z^\star}
\def\wstar{W^\star}

\def\mupmum{\mu^+\mu^-}

\def\rts{\sqrt s}
\def\ie{{\it i.e.}}
\def\eg{{\it e.g.}}

\def\anti{\overline}

\def\wm{W^-}
\def\mw{m_W}
\def\mz{m_Z}
\def\h{h}
\def\mh{m_{\h}}
\def\gamh{\Gamma_{\h}^{\rm tot}}

\def\hsm{h_{SM}}
\def\mhsm{m_{\hsm}}
\def\gamhsm{\Gamma_{\hsm}^{\rm tot}}
\def\tanb{\tan\beta}
\def\hl{h^0}
\def\mhl{m_{\hl}}

\def\ha{A^0}
\def\mha{m_{\ha}}
\def\gamha{\Gamma_{\ha}^{\rm tot}}
\def\hh{H^0}
\def\mhh{m_{\hh}}
\def\gamhh{\Gamma_{\hh}^{\rm tot}}
\def\fbi{~{\rm fb}^{-1}}
\def\fb{~{\rm fb}}
\def\mev{~{\rm MeV}}
\def\gev{~{\rm GeV}}
\def\tev{~{\rm TeV}}
\def\stop{\widetilde t}
\def\mstop{m_{\stop}}
\def\mt{m_t}

\def\gampzero{\Gamma_{\pzero}^{\rm tot}}

\def\tanb{\tan\beta}

\begin{document}
\font\fortssbx=cmssbx10 scaled \magstep2
\hbox to \hsize{
$\vcenter{
\hbox{\fortssbx University of California - Davis}
}$
\hfill
$\vcenter{
\hbox{\bf UCD-98-5} 
\hbox{\bf hep-ph/9802258}
\hbox{February, 1998}
}$
}
\medskip

\title{Physics at a Muon Collider
\thanks{
To appear in Proceedings of {\it Workshop on 
Physics at the First Muon Collider
and at the Front End of a Muon Collider}, Fermilab, Chicago, November 6--9, 
1997, editors S. Geer and R. Raja, AIP Press.
Work supported in part by U.S. Department of Energy 
grant No. DE-FG03-91ER40674 and by the U.C. Davis Institute
for High Energy Physics.}
}
\author{John F. Gunion}
\address{Davis Institute for High Energy Physics,
Department of Physics, University of California,
Davis, CA 95616, USA}

\maketitle

\thispagestyle{empty}

\newlength{\captsize} \let\captsize=\small 

\def\mVV{M_{VV}}
\def\muc{$\mu$C}
\def\ec{$e$C}
\def\ellc{$\ell$C}

\begin{abstract}
I discuss the exciting prospects for exploring a wide range of new physics 
at a low-energy muon collider.
\end{abstract}

The physics possibilities for muon colliders (\muc's) are enormous.
An incomplete list includes:
front-end physics;
$Z$ physics;
Higgs physics, especially $s$-channel factory production;
precision $\mw$, $\mt$ measurements;
deep-inelastic physics, including lepto-quarks and contact interactions;
supersymmetry, including $s$-channel sneutrino production in
R-parity violating models;
strong-$WW$ sector physics; light and heavy technicolor resonances; 
and new $Z^\prime$'s. 
No matter what physics lies beyond the Standard Model,
the muon collider will be a very exciting machine.
In this talk, I will emphasize those topics that are relevant to a first
`low'-energy muon collider ($E_{\rm beam}\sim 50-250\gev$), paying
special attention to $s$-channel resonance probes of new physics.

The instantaneous luminosity, $\call$, possible
for $\mupmum$ collisions depends on $E_{\rm beam}$ and the percentage
Gaussian spread in the beam energy, denoted by $R$. 
The small level of bremsstrahlung and absence of beamstrahlung implies
that very small $R$ can be achieved. The (conservative)
luminosity assumptions for this workshop were:~\footnote{For
yearly integrated luminosities, we use the standard convention
of $\call=10^{32}$cm$^{-2}$s$^{-1}\To L=1\fbi/{\rm yr}$.}
\begin{itemize}
\item
$\call\sim (0.5,1,6) \cdot 10^{31}\cmsi$ for $R=(0.003,0.01,0.1)\%$
at $\rts\sim 100\gev$;
\item
$\call\sim (1,3,7) \cdot 10^{32}\cmsi$,
at $\rts\sim (200,350,400)\gev$, $R\sim 0.1\%$.
\end{itemize}
With modest success in the collider design, at least a factor of 2
better can be anticipated. Note that
for $R\sim 0.003\%$ the Gaussian spread in $\rts$, given by
$\sigrts\sim 2\mev\left({R\over 0.003\%}\right)\left({\rts\over
100\gev}\right)$,  can be comparable to
the few MeV widths of very narrow resonances
such as a light SM-like Higgs boson, sneutrino resonance, or technicolor boson.
This is critical since the effective resonance cross section $\overline\sigma$
is obtained by convoluting a Gaussian $\rts$ distribution of width
$\sigrts$ with the standard $s$-channel Breit Wigner resonance 
cross section $\sigma(\rtshat)=
4\pi\Gamma(\mu\mu)\Gamma(X)/([\shat-M^2]^2+[M\Gamma^{\rm tot}]^2)$.
For $\rts=M$, the result,~\footnote{In actual
numerical calculations, bremsstrahlung smearing is
also included (see Ref.~\cite{bbgh}).}              
\begin{equation}
\small
\overline\sigma\simeq {\pi\sqrt{2\pi} \Gamma(\mu\mu)\, \br( X) \over
M^2\sigrts}
\times \left(1+ {\pi\over 8}\left[{\Gamma^{\rm tot}\over
\sigrts}\right]^2\right)^{-1/2}\,, 
\label{sigmaform}
\end{equation}
will be maximal if $\Gamma^{\rm tot}$ is small and 
$\sigrts\sim\Gamma^{\rm tot}$.~\footnote{Although smaller 
$\sigrts$ (\ie\ smaller $R$) implies smaller $\call$, 
the $\call$'s given earlier are such that when $\Gamma^{\rm tot}$ 
is in the MeV range
it is best to use the smallest $R$ that can be achieved.}
Also critical to scanning a narrow resonance 
and for precision $\mw$ and $\mt$ measurements is the ability~\cite{raja}
to tune the beam energy to one part in $10^{6}$.
Finally, by constructing the muon collider
at a facility (such as Fermilab) with a high energy proton beam one
opens up the possibility of having a $\mu p$ collider option. 
The luminosity expected for $200\gev$ 
$\mu^+$ and $\mu^-$ beams in collision with the
$1\tev$ proton beam of the Tevatron (yielding $\rts=894\gev$) is $\call\sim 
1.3\cdot 10^{33}\cmsi$.

\section*{Physics}
\bit
\item
{\bf\boldmath Front-End and $\mu$ Beam Physics}
\eit
A proton driver and intense cooled low-energy muon beam
will be the first components of the muon collider to be constructed.
These alone will yield a large program of ``front-end'' physics. In particular,
low-energy hadronic physics ($p,\anti p,K,\pi$) \cite{littenberg} and
low-energy neutrino physics (analogous to the LSND and BOONE experiments)
can be explored with much improved statistics \cite{kayser}. Great 
strides in stopped/slow intense muon beam physics 
(\eg\ $g_\mu-2$, $\mu N\to eN$ conversion, $\mu\to eee$, $\mu\to e\gam$)
will also be possible  \cite{marciano,molzon}. The search for 
$\mu N\to eN$ deserves special mention as it would probe
for lepton-flavor violation at a level that is generically expected
from any one of several sources present in supersymmetric models
and other extensions of the SM \cite{marciano,raby}. 
\bit
\item
{\bf\boldmath $Z$ Physics}
\eit
A low-energy muon collider could be run as a $Z$ factory that would 
quickly exceed statistical levels achieved at LEP and SLC/SLD. Using
$\sigma(Z)_{\rm peak}
\sim 6\times 10^7 \fb$ ($\Gamma_Z^{\rm tot}\gg\sigrts$) and
$\call\sim 10^{32}\cmsi$ for the \muc\ (assuming
$R\gsim 0.1\%$ as is perfectly acceptable for $Z$ physics)
leads to $\sim 6\cdot 10^7$ $Z$'s per year (about four times the
best yearly rate achieved at LEP); partial ($\sim 20\%$) 
polarization for {\it both} beams would be 
automatic.~\footnote{At the \muc,
substantial polarization ($\gsim 50\%$) for both beams 
can be achieved only with a significant sacrifice in luminosity.}

The many important physics topics include the following.
(a) 
$B_s-\anti B_s$ mixing. 
(b) 
An improved measurement of 
$\sin^2\theta_W^{\rm eff}$,
as probed via $A_{LR}$ or $A_{FB}$, to resolve
the LEP/SLD disagreement.
(c) 
Improved $\alpha_s$ determination.
(d) 
CP Violation, as probed \eg\  by $Z\to B_d\anti B_d$ 
($B_d,\anti B_d \to\psi K_S$) decays.
(e) 
$\tau$ Michel parameters using $Z\to\tauptaum$ decays.
(f)
Separation of color-octet from  color-singlet $J/\psi$ production;
detailed distributions in the final state would allow this, but LEP
statistics have proved inadequate.
(g)
Improved limits (or actual observation) of flavor-changing-neutral-current
(FCNC) rare decays; the current limits on
$Z\to e\mu$, $Z\to e\tau$, $Z\to \mu\tau$ from the PDG \cite{pdg} are
$1.7\times 10^{-6}$, $9.8\times 10^{-6}$, $1.7\times 10^{-5}$,
respectively. Some types of new physics would predict such decays at
levels just below this.
(h)
Improved limits on or observation of
$Z\to \gam X$ decays, which probe many kinds of new physics. 

Of these, (a) (b) and (c) received attention during the workshop
\cite{demarteau}. 
With the expected $L\sim 1\fbi/{\rm yr}$ (20\% polarization for the beams 
being acceptable) one can achieve $\Delta\alpha_s\sim 0.001$
(vs. the current $\sim 0.003$) and an actual measurement of
the $x_s$ parameter of $B_s-\anti B_s$ mixing (for which LEP
provides only an upper bound). Using $\Delta A_{LR}=({\cal P}\sqrt N)^{-1}$,
where ${\cal P}={P^+-P^-\over 1-P^+P^-}$,
and $\Delta\sin^2\theta_{\rm eff}^{\rm lept}\sim \Delta A_{LR}/7.9$,
one finds $\Delta\sin^2\theta_{\rm eff}^{\rm lept}\sim 0.0001$ (current
error being $\lsim 0.00025$ from combined LEP data) for
a sample of $\sim 10^7$ $Z$'s 
with $P^\pm\sim \pm 30\%$ polarization for the $\mu^\pm$ beams. 
This would take at most a few 
years of operation for current \muc\ designs.

The list of new physics probed by $Z\to\gam X$ decays is impressive.
The factor of ten improvement in sensitivity
to such decays, coming from the $\gsim 10^8$ $Z$'s 
produced after a few years at a 
muon collider $Z$ factory, would be very valuable.
(i) An anomalous $ZZ\gam$ CP-conserving and/or
CP-violating coupling that might arise beyond the SM would lead to
$Z\to \gam Z^*\to \gam \nu\anti\nu$ events; current limits
from LEP \cite{l3zgam} and D0 \cite{d0zgam} are already constraining
on SM extensions.
(ii) Anomalous trilinear and quartic couplings can lead to
$Z\to \gam\gam\gam$ events. The SM prediction is 
$\br(\gam\gam\gam)\sim 10^{-9}$ while the
current limit is $\lsim 10^{-5}$;  many SM extensions
predict branching ratios of this latter size  \cite{renard}. 
(iii) The magnitude of the $\nu_\tau$ magnetic moment 
is very relevant to understanding basic neutrino properties and 
can have a large impact on predictions for this source of dark matter.
Non-zero $\mu_{\nu}$ leads to $\gam$ radiation from the final $\nu$ 
and $\anti\nu$ in $Z\to \nu\anti\nu$. 
Current LEP data yields \cite{l3zgam} $\mu_{\nu_\tau}\lsim 3.3\times
10^{-6}\mu_{\rm B}$ (90\% CL).  Limits from elsewhere are competitive.
(iv) Improved limits on axions would be possible from searches for
$Z\to\gam A$, where $A$ decays invisibly. Current 
limits on this branching ratio
from LEP are \cite{l3zgam} ${\rm few}\times 10^{-6}$.  
If axions exist, $Z\to\gam A$
decays might be observed with improved sensitivity. Stronger limits
would significantly constrain many models.
(v) Also of interest are decays of the type $Z\to \gam+{\rm meson}$, 
\eg\ $Z\to \gam\pi^0,\gam\eta,\gam\,J/\psi,\ldots$  Current limits on
such branching ratios are $\lsim {\rm few}\times 10^{-5}$ \cite{pdg}.
Not only has there been much dispute about the SM predictions, but also
new physics could enter. Surprises could emerge with any increase in
sensitivity.
(vi) 
A particularly important probe of technicolor theories
is the $Z\to\gam\gam\gam,\gam\ell^+\ell^-,\gam \etmiss,\gam q\anti q,\gam gg$ 
class of decays expected from $Z\to \gam \pzero$, where $\pzero$
is an electrically neutral pseudo-Nambu-Goldstone boson (PNGB) that can
decay to one or more of the indicated channels \cite{simmons}.
The predicted branching ratio for $Z\to \gam \pzero$ is
$\br(Z\to\gam \pzero)\sim 10^{-5}\left({123\gev\over f}\right)^2
(N_{TC}A_{Z\gam})^2\beta^3$,
where the anomaly factor $A_{Z\gam}$ is $\calo (.05-1)$
and $f$ is the technipion decay constant. Improving
limits in the above channels by a factor of ten would rule out light 
$\pzero$'s in many technicolor models, whereas 
currently the light PNGB's of most models would have escaped detection.
(vii) 
Finally, we note that many of the above exotic decays could have
large branching ratio if the particles involved are composite.

Overall, 
a muon collider $Z$ factory would have the luminosity needed to resolve
some important outstanding $Z$ physics and would provide increased
sensitivity to very important rare processes that probe new physics.
\bit
\item
{\bf Higgs Physics}
\eit
The potential of the muon collider for Higgs physics is truly outstanding.
First, it should be emphasized that 
away from the $s$-channel Higgs pole, $\mupmum$ and $\epem$ colliders
have similar capabilities for the same $\rts$ and $\call$ (barring
unexpected detector backgrounds at the muon collider). At $\rts=500\gev$,
the design goal for a $\epem$ linear collider (\ec) is $L=50\fbi$ per year.
The conservative $\call$
estimates given earlier suggest that at $\rts=500\gev$
the \muc\ will accumulate {\it at least} $L=10\fbi$ per year.  If this can
be improved somewhat, the \muc\ would be fully competitive with the \ec.
We will use the notation of \ellc\ for either a \ec\ or \muc\ operating
at moderate to high $\rts$.

\begin{figure}[h]
\begin{center}
{\epsfig{file=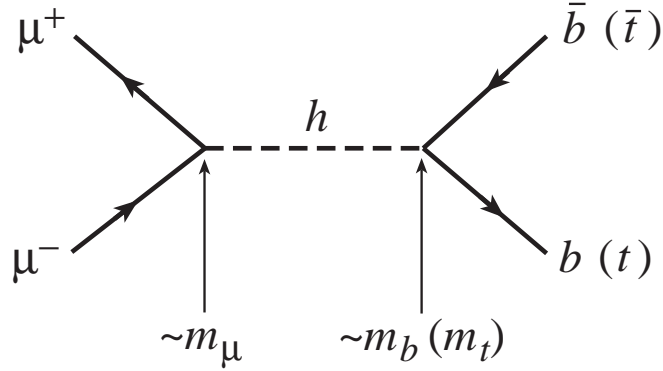,width=3.5in}}
\caption{{
Feynman diagram for $s$-channel production of a Higgs boson.\label{schanfig}}}
\end{center}
\end{figure}

\begin{figure}[h]
\begin{center}
{\epsfig{file=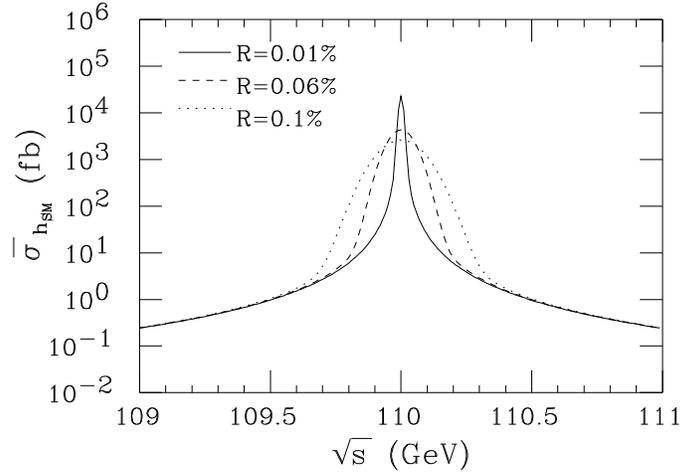,width=3.5in}}
\caption{ The effective cross section, $\sighsmbar$,
for $R=0.01\%$, $R=0.06\%$, and $R=0.1\%$ vs. $\protect\rts$ for 
$\mhsm=110\gev$.
\label{gausssigma}}
\end{center}
\end{figure}

Of course, the totally unique feature of the \muc\ is the very large
cross section expected for production of a Higgs boson in the $s$-channel
when $\rts=\mh$, see Fig.~\ref{schanfig} \cite{bbgh}. Small $R$
is crucial as it leads to
dramatically increased peaking of $\sighbar$ [\Eq{sigmaform}] at $\rts\sim\mh$,
as illustrated in Fig.~\ref{gausssigma} 
for a SM Higgs ($\hsm$) with $\mhsm=110\gev$ ($\gamhsm\sim 3\mev$).

\medskip
\noindent{\bf A Standard Model-Like Higgs Boson}
\medskip

For SM-like $\h\to WW,ZZ$ couplings, $\gamh$ becomes big
if $\mh\gsim 2\mw$, and 
$\sighbar\propto \br(\h\to\mupmum)$  [\Eq{sigmaform}] will
be small; $s$-channel production will not be useful.
But, as shown in Fig.~\ref{gausssigma},
$\sighbar$ is enormous for small $R$ when the $\h$ is light, as is
very relevant in supersymmetric models where the light SM-like $\hl$ has
$\mhl\lsim 150\gev$.
In order to make use of this large cross section, we must
first center on $\rts\sim \mh$. Once this is done we proceed to
the precision measurement of the Higgs boson's properties.

\begin{figure}[h]
\begin{center}
{\epsfig{file=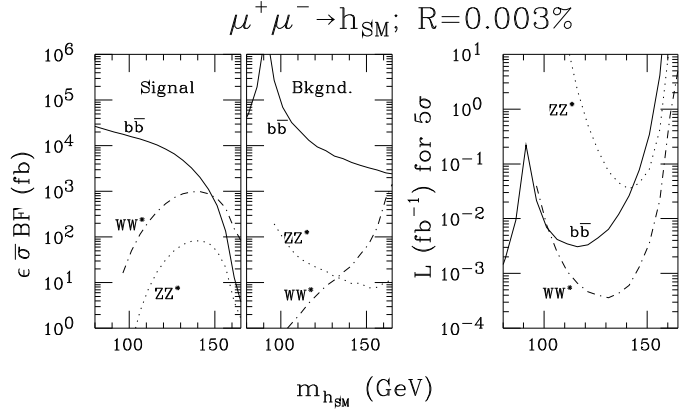,width=3.5in}}
\caption{{
SM rates and $L$ required for $5\sigma$ observation as
a function of $\mhsm$, for $R=0.003\%$.\label{smr003}}}
\end{center}
\end{figure}

For a SM-like Higgs with $\mh\lsim 2\mw$ one expects~\cite{jfghiggs}
$\Delta\mh\sim 100\mev$ from LHC data ($L=300\fbi$) (smaller if \ellc\
data is available). Thus, a final ring that
is fully optimized for $\rts\sim\mh$ can be built.
Once it is operating, we scan over the appropriate $\Delta\mh$ interval so as
to center on $\rts\simeq\mh$ within a fraction of $\sigrts$.
For $\mh$ of order 100 GeV, $R=0.003\%$ implies $\sigrts\sim 2\mev$.
The luminosity required for a $5\sigma$ observation of the SM Higgs boson
with $\rts=\mhsm$ is plotted (along with individual signal and background
rates) in Fig.~\ref{smr003}. In the ``typical'' $\mh\sim 110\gev$ case,
$\Delta\mh\sim 100\mev$ implies that $\Delta\mh/\sigrts \sim 50$ 
points are needed to center within $\lsim \sigrts$.  From Fig.~\ref{smr003}
we find that each point requires $L\sim 0.0015\fbi$ in order to observe
or eliminate the $\h$ at the $3\sigma$ level, implying a total of 
$\ltot\leq 0.075\fbi$ is needed for centering. Thus, for 
the anticipated $L\sim 0.05-0.1\fbi/{\rm yr}$, centering would take
no more than a year. However, 
for $\mh\simeq\mz$ a factor of 50 more $\ltot$ is required just for centering
because of the large $Z\to b\anti b$ background.
Thus, for the anticipated $\call$ the \muc\ is not useful if 
the Higgs boson mass is too close to $\mz$.

Once centered, we will wish to measure with precision:
(i) the very tiny Higgs width ---
$\gamh=1-10\mev$ for a SM-like Higgs with $\mh\lsim 140\gev$;
(ii) $\sigma(\mupmum\to \h\to X)$ for 
$X=\tauptaum, b\anti b,c\anti c,W\wstar,Z\zstar$.                  
The accuracy achievable was studied in Ref.~\cite{bbgh}. The three-point
scan of the Higgs resonance described there is the optimal procedure for
performing both measurements simultaneously. We summarize
the resulting statistical
errors in the case of a SM-like $\h$ with $\mh=110\gev$, assuming $R=0.003\%$
and an integrated (4 to 5 year) $\ltot=0.4\fbi$.~\footnote{For $\sigma\br$
measurements, $\ltot$ devoted to the optimized three-point scan 
is equivalent to $\sim\ltot/2$ at the $\rts=\mh$ peak.}
One finds $1\sigma$ errors for $\sigma\br(X)$ of $8,3,22,15,190\%$
for the $X=\tauptaum,b\anti b, c\anti c,W\wstar,Z\zstar$ channels,
respectively, and a $\gamh$ error of 16\%. These results assume
the $\tau,b,c$ tagging efficiencies described in Ref.~\cite{bking}.
We now consider how useful measurements at these accuracy
levels will be.

If only $s$-channel Higgs factory \muc\ data are available
(\ie\ no $Z\h$ data from an \ec\ or \muc), then the 
$\sigma\br$ ratios (equivalently squared-coupling ratios~\footnote{From
\Eq{sigmaform}, $\sigma(\mupmum\to\h\to X)$ provides
a determination of $\Gamma(\h\to\mupmum)\br(\h\to X)$ (which
is proportional to the $(X\h)^2$ squared coupling) when $\sigrts\gsim
\gamh$, as is the case.}) that will be most effective for 
discriminating between the SM Higgs
boson and a SM-like Higgs boson such as the $\hl$ of supersymmetry are
${(W\wstar\h)^2\over (b\anti b\h)^2},$
${(c\anti c\h)^2\over (b\anti b\h)^2},$
${(W\wstar\h)^2\over (\tauptaum\h)^2},$ and
${(c\anti c\h)^2\over (\tauptaum\h)^2}.$
The $1\sigma$ errors (assuming $\ltot=0.4\fbi$ at $\mh=110\gev$)
for these four ratios are $15\%$, $20\%$, $18\%$ and $22\%$, respectively.
Systematic  errors for $(c\anti c\h)^2$ and $(b\anti b\h)^2$
of order $5\%-10\%$
from uncertainty in the $c$ and $b$ quark mass will also enter.
In order to interpret these errors one must compute
the amount by which the above ratios differ in the minimal
supersymmetric model (MSSM) vs. the SM for $\mhl=\mhsm$. The percentage
difference turns out to be essentially identical for
all the above ratios and is a function almost only of the MSSM Higgs sector
parameter $\mha$, with very little dependence on $\tanb$ or top-squark
mixing. At $\mha=250\gev$ ($420\gev$) one finds MSSM/SM $\sim 0.5$ ($\sim
0.8$). Combining the four independent ratio measurements 
and including the systematic errors, one concludes that 
a $>2\sigma$ deviation from the SM predictions would
be found if the observed Higgs is the MSSM $\hl$ and $\mha < 400\gev$.
Note that the magnitude of the deviation would provide
a determination of $\mha$.

\begin{figure}[h]
\begin{center}
{\epsfig{file=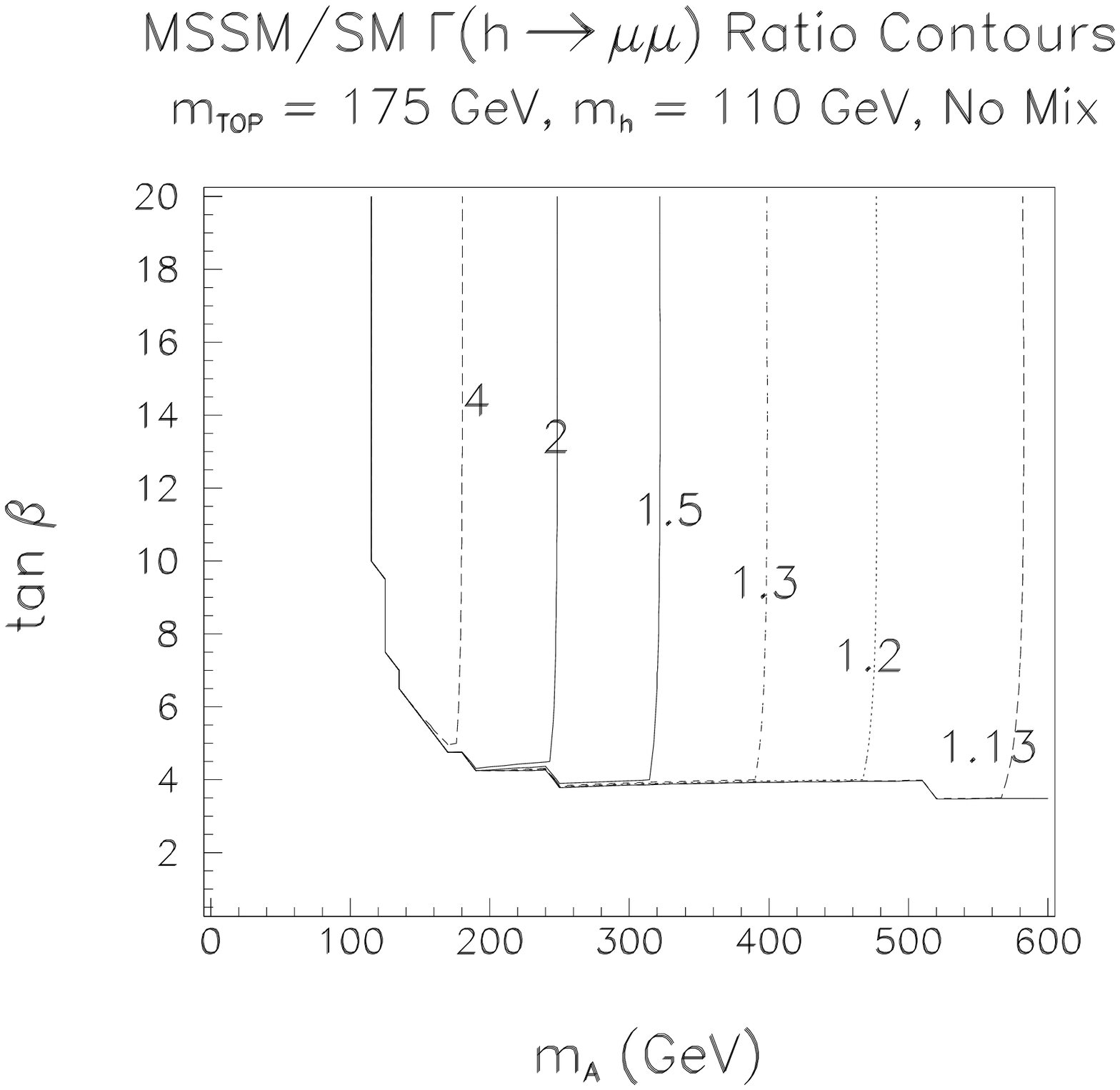,width=3.5in}}
\caption{We give $(\mha,\tanb)$ parameter space contours
for ${\Gamma(\hl\to\mupmum)\over \Gamma(\hsm\to\mupmum)}$:
no-squark-mixing, $\mhl,\mhsm=110\gev$. \label{mumucontours}}
\end{center}
\end{figure}

If, in addition to the $s$-channel measurements we also have \ellc\
$\rts=500\gev$, $\ltot=200\fbi$ data, it will be possible
to discriminate at an even more accurate level between the $\hl$
and the $\hsm$.  The most powerful technique for doing so employs
the four determinations of $\Gamma(\h\to\mupmum)$ below:
\bea
&{[\Gamma(\h\to\mupmum)\br(\h\to
b\anti b)]_{\mu{\rm C}}\over \br(\h\to b\anti b)_{\ell{\rm C}}};
~~~~~~~~~~~~~~~~
{[\Gamma(\h\to\mupmum)\br(\h\to
W\wstar)]_{\mu{\rm C}}\over\br(\h\to W\wstar)_{\ell{\rm C}}};&\nn\\
&{[\Gamma(\h\to\mupmum)\br(\h\to
Z\zstar)]_{\mu{\rm C}}[\gamh]_{\mu{\rm C}+\ell{\rm C}}
\over\Gamma(\h\to Z\zstar)_{\ell{\rm C}}};
~~~
{[\Gamma(\h\to\mupmum)\br(\h\to
W\wstar)\gamh]_{\mu{\rm C}}\over\Gamma(\h\to W\wstar)_{\ell{\rm C}}}.&
\label{gammumu}
\eea
The resulting $1\sigma$ error for $\Gamma(\h\to\mupmum)$ is $\lsim 5\%$.
Fig.~\ref{mumucontours}, which plots the ratio of
the $\hl$ to $\hsm$ partial width in $(\mha,\tanb)$ parameter
space for $\mhl=\mhsm=110\gev$, shows that this level of error allows one
to distinguish between the $\hl$ and $\hsm$ at the $3\sigma$ level 
out to $\mha\gsim 600\gev$. This result holds for all $\mh \lsim 2\mw$
($\mh\neq\mz$).
Additional advantages of a $\Gamma(\h\to\mupmum)$ measurement are:
(i) there are no systematic uncertainties arising from uncertainty
in the muon mass; (ii) the error on
$\Gamma(\h\to\mupmum)$ increases only very slowly as the $s$-channel
$\ltot$ decreases,~\footnote{This is because
the $\Gamma(\h\to\mupmum)$ error is dominated by the $\rts=500\gev$
measurement errors.} in contrast to the errors for the previously
discussed ratios of branching ratios from the \muc\
$s$-channel data which scale as $1/\sqrt{\ltot}$.
Finally, we note that $\gamh$ alone cannot be used
to distinguish between the MSSM and SM in model-independent way. 
Not only is the error substantial ($\sim 12\%$ if we combine \muc, $L=0.4\fbi$ 
$s$-channel data with \ellc, $L=200\fbi$ data)
but also $\gamh$ depends on many things, including (in the MSSM) the
squark-mixing model. Still, deviations from SM predictions are generally
substantial if $\mha\lsim 500\gev$.

Precise measurements of the couplings of the SM-like Higgs boson could
reveal many other types of new physics. For example, if a significant
fraction of a fermion's mass is generated radiatively (as opposed
to arising at tree-level), then the $\h f\anti f$ coupling 
and associated partial width will deviate from SM expectations
\cite{borzumati}. Deviations of order 5\% to 10\% (or more)
in $\Gamma(\h\to \mupmum)$
are quite possible and, as discussed above, potentially detectable.

\medskip
\noindent{\bf\boldmath The MSSM $\hh$, $\ha$ and $\hpm$}
\medskip

We begin by recalling \cite{jfghiggs} that
the possibilities for $\hh,\ha$ discovery are limited at other machines.
(i) Discovery of $\hh,\ha$ is not possible at LHC
for all $(\mha,\tanb)$: \eg\ if $\mstop=1\tev$, consistency
with the observed value of $\br(b\to s\gam)$
requires $\mha>350\gev$, in which case the LHC will
not detect the $\hh,\ha$ 
if $\tanb\gsim 3$ (and below a much higher $\mha$-dependent
value).
(ii) At $\rts=500\gev$, $\epem\to \hh\ha$ 
pair production probes only to $\mha\sim\mhh\lsim 230-240\gev$.
(iii)
A $\gam\gam$ collider could potentially probe up
to $\mha\sim\mhh\sim 0.8 \rts\sim 400\gev$, but only
for $\ltot\gsim 150-200\fbi$.

Thus, it is noteworthy that 
$\mupmum\to\hh,\ha$ in the $s$-channel 
potentially allows production and study
of the $\hh,\ha$ up to $\mha\sim\mhh\lsim \rts$. 
To assess the potential, let us (optimistically) 
assume that a total of $\ltot=50\fbi$ (5 yrs
running at $<\call>=1\times 10^{33}$) can be accumulated for 
$\rts$ in the $ 250-500\gev$ range. (We note that $\gamha$ and $\gamhh$,
although not big, are of a size such that resolution of $R\gsim 0.1\%$
will be adequate to maximize the $s$-channel cross section, thus
allowing for substantial $\call$.)

\begin{figure}[h]
\begin{center}
{\epsfig{file=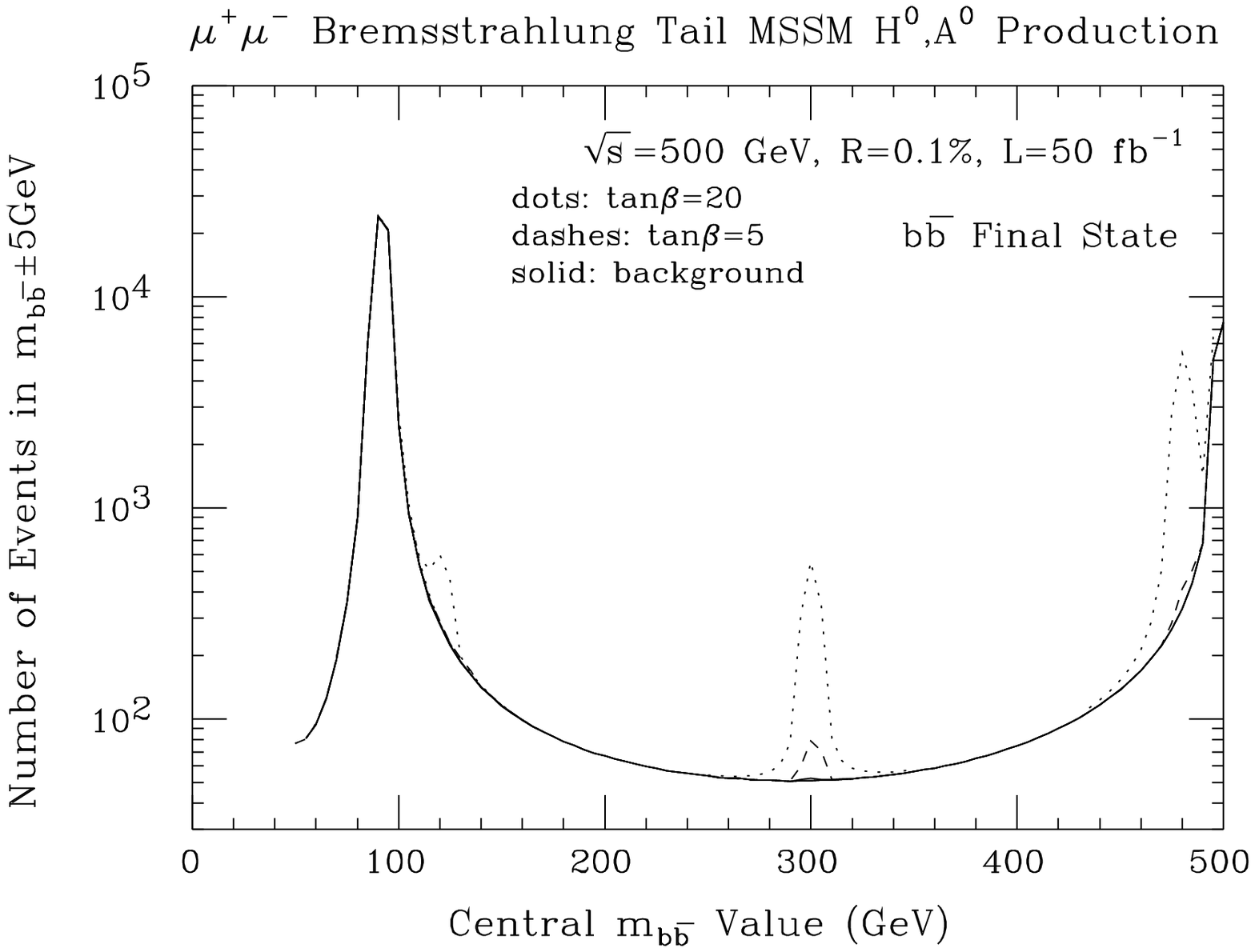,width=3.5in}} 
\caption{ $N(b\anti b)$ in the $\mbb\pm5\gev$ interval
vs. $\mbb$ for $\protect\rts=500\gev$, $\ltot=50\fbi$,
and $R=0.1\%$: peaks are shown for
$\mha=120$, $300$ or $480\gev$, with $\tanb=5$ and $20$ in each case.
 \label{mbbscan}}
\end{center}
\end{figure}

There are then several possible scenarios.
(a) If we have some preknowledge or restrictions
on $\mha$ from LHC discovery or 
from $s$-channel measurements of $\hl$ properties,
then $\mupmum\to\hh$ and $\mupmum\to\ha$ can be studied with
precision for all $\tanb\gsim 1-2$.
(b) 
If we have no knowledge of $\mha$ other than $\mha\gsim 250-300\gev$
from LHC, then we might wish to search for the $\ha,\hh$
in $\mupmum\to\hh,\ha$ by scanning over $\rts=250-500\gev$.
If their masses lie in this mass range, then their discovery
by scanning will be possible for most of $(\mha,\tanb)$ parameter
space such that they cannot be discovered at the LHC (in particular,
if $\mha\gsim 250\gev$ and $\tanb\gsim 4-5$).
(c) Alternatively, if the \muc\
is simply run at $\rts=500\gev$ and $\ltot\sim 50\fbi$ is accumulated,
then $\hh,\ha$ in the $250-500\gev$ mass range
can be discovered in the $\rts$ bremsstrahlung tail if the $b\anti b$
mass resolution (either by direct reconstruction or
hard photon recoil) is of order $\pm 5\gev$ and if $\tanb\gsim 6-7$ (depending
on $\mha$). Typical peaks are illustrated in 
Fig.~\ref{mbbscan}.~\footnote{SUSY decays are assumed to be absent in
this and the following figure.}

\begin{figure}[h]
\begin{center}
{\epsfig{file=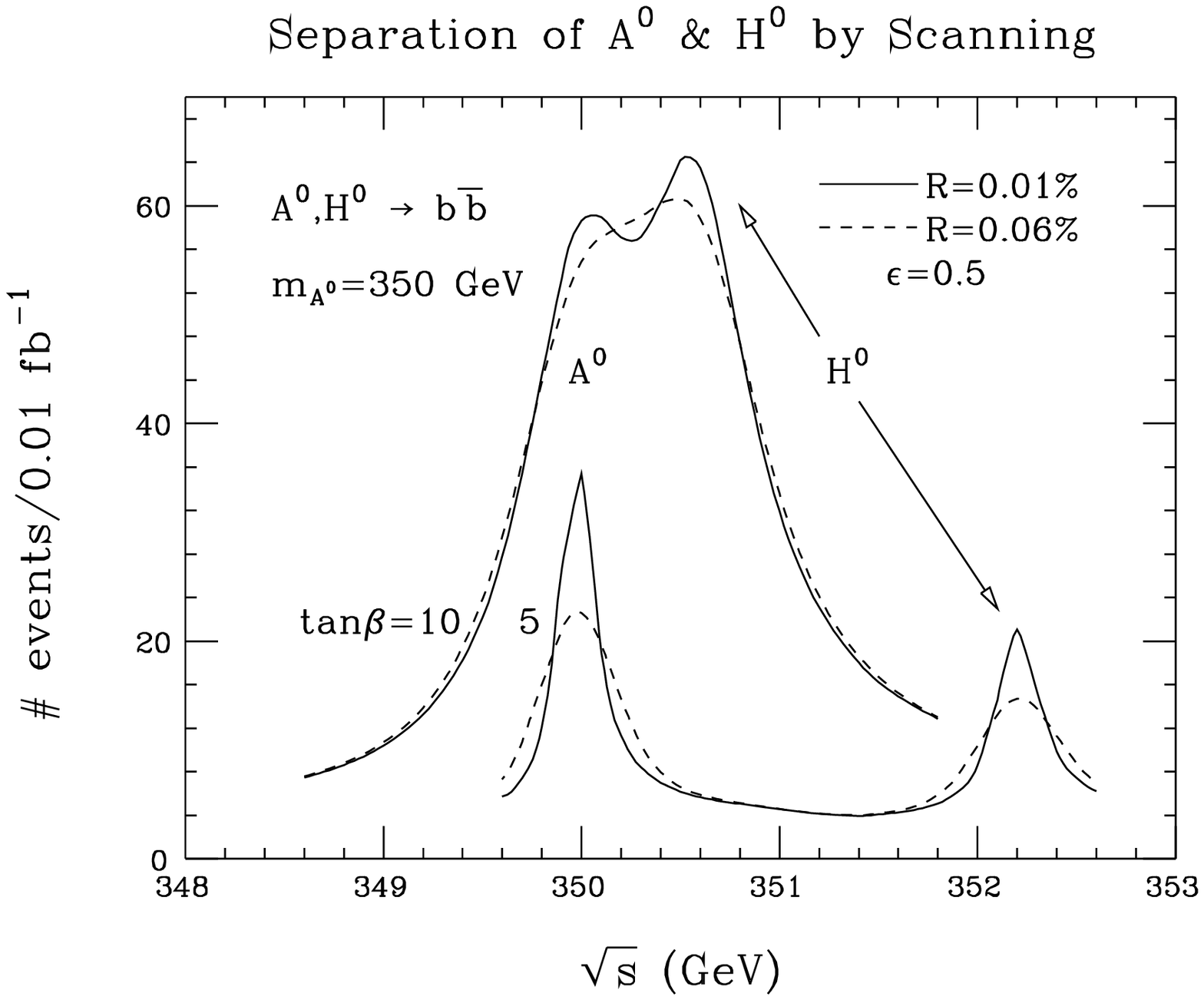,width=3.5in}}
\caption{ $N(b\anti b)$ (for $0.01\fbi$)
vs. $\protect\rts$, for $\mha=350\gev$ $\hh,\ha$ resonance (with
$\tanb=5$ and 10), including the $b\anti b$ continuum background.
\label{hhhasusyrtsscan}}
\end{center}
\end{figure}

Finally, once the closely degenerate $\ha,\hh$ are discovered,
it will be extremely interesting to be able to separate the resonance
peaks.  This will probably only be possible at a muon collider
with small $R\lsim 0.01\%$ if $\tanb$ is large, as illustrated in
Fig.~\ref{hhhasusyrtsscan}.

We end with just a few remarks on the possibilities for production
of $\hh\ha$ and $\hp\hm$ pairs at a high energy \muc\ (or \ec).
Since $\mha\gsim 1\tev$ cannot be ruled out simply on the basis
of hierarchy and naturalness (although fine-tuning is stretched),
it is possible that energies of $\rts >2\tev$ could be
required for pair production.
If available, then it has been shown \cite{gk,fengmoroi} that
discovery of $\hh\ha$ in their $b\anti b$ or $t\anti t$ 
decay modes and $\hp\hm$ in their $t\anti b$ and $b\anti t$ decays
will be easy for expected luminosities, even if SUSY decays are
present. As a by-product, the masses will be measured
with reasonable accuracy. 

Regardless of whether we see the $\hh,\ha$ in $s$-channel production
or via pair production, one can measure
branching ratios to other channels, including supersymmetric pair
decay channels with good accuracy.
In fact, the ratios of branching ratios and the value
of $\mha\sim\mhh\sim\mhpm$ will be measured with sufficient
accuracy that, in combination with one gaugino mass, say the chargino
mass (which will also presumably be well-measured)
it will be possible \cite{gk} to discriminate 
with incredible statistical significance between
different closely similar GUT scenarios for the 
GUT-scale soft-supersymmetry-breaking masses.
Thus, Higgs pair production could be very valuable in the
ultimate goal of determining all the soft-SUSY-breaking parameters.

Finally, entirely unexpected decays of the heavy Higgs bosons of SUSY
(or other extended Higgs sector) could be present. For example,
non-negligible branching ratios for $\hh,\ha\to t\anti c+c\anti t$ FCNC 
decays are not inconsistent with current theoretical model-building ideas
and existing constraints \cite{reina}. The muon collider $s$-channel 
$\mupmum\to \hh,\ha$ event rate is sufficient to probe rather small values
for such FCNC branching ratios.

\medskip
\noindent{\bf Exotic Higgs Bosons}
\medskip

If there are doubly-charged Higgs bosons, $\em\em\to \dmm$
probes $\lam_{ee}$ and $\mu^-\mu^-\to \dmm$ probes $\lam_{\mu\mu}$,
where the $\lam$'s are the strengths of the Majorana-like couplings
\cite{jfgemem,frampton,cuypers}.  
Current $\lam_{ee,\mu\mu}$ 
limits are such that factory-like production of a $\dmm$ 
is possible if $\gamdmm$ is small.
Further, a $\dmm$ with $\mdmm\lsim 500-1000\gev$
will be seen previously at the LHC 
(for $\mdmm\lsim 200-250\gev$ at TeV33) \cite{kpitts}.
For small $\lam_{ee,\mu\mu,\tau\tau}$ in the range that
would be appropriate, for example, for the $\dmm_L$ 
in the left-right symmetric model see-saw
neutrino mass generation context, it may be that
$\gamdmm\ll\sigrts$,~\footnote{For small $\lam_{ee,\mu\mu,\tau\tau}$,
$\gamdmm$ is very small
if the $\dmm\to\wm\wm$ coupling strength is very small or zero, as required
to avoid naturalness problems for $\rho=\mw^2/[\cos^2\theta_w\mz]^2$.}
leading to
$\overline\sigma_{\ell^-\ell^-\to\dmm}\propto \lam_{\ell\ell}^2/\sigrts$.
Note that the absolute rate for $\ell^-\ell^-\to\dmm$
yields a direct determination
of $\lam_{\ell\ell}^2$, which, for a $\dmm$ with very small $\gamdmm$, will
be impossible to determine by any other means. The relative branching
ratios for $\dmm\to e^-e^-,\mu^-\mu^-,\tau^-\tau^-$ will then yield values
for the remaining $\lam_{\ell\ell}^2$'s.
Because of the very small $R=0.003\%-0.01\%$ achievable at a muon collider,
$\mu^-\mu^-$ collisions will probe much weaker $\lam_{\mu\mu}$ coupling
than the $\lam_{ee}$ coupling that can be probed in $e^-e^-$ collisions.
In addition, it is natural to anticipate that $\lam_{\mu\mu}^2\gg\lam_{ee}^2$.

\bit
\item
{\bf\boldmath Precision Measurements of $\mw$ and $\mt$}
\eit
Let us consider the extent to which the muon collider
could contribute to precision measurements of $\mw$ and $\mt$.
Current expectations for the Tevatron, LHC and \ec\ 
for various benchmark accumulated luminosities appear in
Table~\ref{fixedl}~\cite{mwmtsnowmass}. 
Note that more than $\ltot=50\fbi$ is not 
useful for these measurements at an electron collider 
since errors become systematics dominated.

At the \muc, $WW$ threshold and $t\anti t $ threshold measurements are the most
accurate ways to determine $\mw$ and $\mt$. Because of the small $R$
and precise beam energy determination possible at the muon collider,
the errors at the \muc\ (given in Table~\ref{fixedl}) \cite{bbghmwmt,berger}
are always statistics dominated and the 
accuracy that can be achieved at the \muc\ is about a factor of
two better that at the \ec\ for the same $\ltot$. However,
lower yearly $\ltot$ is expected at the \muc\ than at the \ec.
Taking $R\sim 0.1\%$ (better is not useful), 
the conservative luminosities given earlier imply 
$\ltot(\rts=2\mw)\sim 1\fbi/{\rm yr}$ and $\ltot(\rts=2\mt)\sim
3\fbi/{\rm yr}$. Based on these inputs, the conclusion seems to be that 
systematics from beam energy spread \etc\ are
low enough at the \muc\ that accuracies for $\mt$ competitive to 1 year
of \ec\ operation can be achieved after 2 years of \muc\ running at a
luminosity that is a factor of two better than
the conservative assumption of the workshop. The \muc\ measurement
of $\mw$ would be competitive only if $\call$ at $\rts\sim 2\mw$ can be much
larger (a factor of ten or so) than current expectations.
Given this, and the fact 
that the precision electroweak determination of $\mhsm$ is
optimized for $\Delta\mw/\Delta\mt\sim 0.02$, it would seem best
to focus on $t\anti t$ threshold measurments at the \muc.

\begin{table}[h]
\caption[fake]{\baselineskip 0pt Comparison of the
achievable precision in $\mw$ and $\mt$ measurements at
different future colliders for different $\ltot$. \label{fixedl}}
\begin{center}
\begin{tabular}{|c|c|c|c|c|c|c|c|c|c|c|}
\hline
 & \multicolumn{2}{c|}{\quad Tevatron}& LHC& $e$C&
\multicolumn{4}{c|}{\quad $\mu$C} \\ \hline
$\ltot$ (fb$^{-1}$)& 2& 10& 10& 50& 1& 3& 10& 50\\ \hline
$\Delta \mw$ (MeV)& 22--35& 11--20& 15& 15--20& 63& 36& 20& 10\\ \hline
$\Delta \mt$ (GeV)& 4&  2& 2& 0.12--0.2& 0.63& 0.36& 0.2& 0.1 \\ \hline
\end{tabular}
\end{center}
\end{table}

Focusing on the $t\anti t$ threshold is further motivated by the fact
that for small $R$ such measurements
are valuable for determining $\alpha_s$, $\Gamma_t^{\rm tot}$ and $|V_{tb}|^2$,
as well as $\mt$. (There is also dependence on $\mhsm$.) A much more
detailed scan is needed to determine these other quantities 
than the two- or three-point scan optimized for just the $\mt$ measurement. 
To give one example, by devoting $\ltot=10\fbi$ to a ten-point scan, one can 
achieve $\Delta \mt\sim 70\mev$ and 
$\Delta\alpha_s\sim 0.0015$ \cite{bbghmwmt}.
\bit
\item
{\bf Supersymmetry}
\eit
The enormous opportunities in this area are detailed in \cite{carena}.
The program that has been developed for linear $\epem$ colliders
is largely applicable also at a $\mupmum$ collider. Discovery of
pair production of supersymmetric particles with pair mass below $\rts$
is generally straightforward, and detailed measurements of their
masses and other properties will generally be possible.
For example, the lepton and/or jet spectrum end points will typically
allow measurement of the LSP mass, the lightest chargino mass,
and at least some slepton masses \cite{paige,lykken,inprep}.
The only drawback at a \muc\ is the loss of luminosity associated
with the large beam polarization(s) that would be useful for
some SUSY studies.
If some of the supersymmetric particles are very heavy, then
the fact that a \muc\ may be able to reach to higher energy than
an \ec\ could ultimately become crucial. Studies \cite{kelly,inprep}
suggest that a very high energy \muc\ operating with high luminosity 
will be able to pin down the GUT-scale boundary conditions of the SUSY
model with considerable precision, despite the fact that many
different types of SUSY particle pairs will be produced.
\bit
\item
{\bf \boldmath $\mu p$ Collisions}
\eit
We consider colliding one of the \muc\ beams ($\mu^+$ or $\mu^-$) with
whatever proton beam is available, \eg\ the $1\tev$ ($820\gev$) $p$ beam
at Fermilab (DESY). Useful benchmark 
possibilities are $E_\mu\otimes E_p=30\gev\otimes 820\gev$ ($\rts=314\gev$),
$50\gev\otimes 1\tev$ ($\rts=447\gev$), and
$200\gev\otimes 1\tev$ ($\rts=894\gev$).
For the $\rts=314\gev$ machine we assume $L=0.1\fbi/{\rm yr}$ 
so as to provide a direct comparison to $ep$ collisions at HERA.
For the Tevatron machines, we assume $L=2$ and $13\fbi/{\rm yr}$,
respectively.~\footnote{The former is the result obtained using 
scaling \cite{palmer} of $\call\propto E_\mu^{4/3}$ starting with
the workshop assumption of
$L=13\fbi/{\rm yr}$ at the $200\gev\otimes 1\tev$ machine.}

As discussed in \cite{ritz}, the $\rts=894\gev$ machine with
$L=13\fbi/{\rm yr}$ yields a big increase (compared
to HERA) in the kinematic limits accessible, allowing exploration
of very large $Q^2$ values at moderate to high $x$ values.
However, event kinematics 
and detector considerations imply that low-$x$ studies would be very difficult.
Backgrounds could be an issue in some kinematic regions.

\medskip
\noindent{\bf Contact and/or Lepto-quark Interactions}
\medskip

The potential of the $\mu p$ collider program is perhaps best illustrated
in the context of contact or lepto-quark ($L_q$) interactions.
The relevant contact interactions are denoted $\Lambda^{\mu q}_{LL,LR,RL,RR}$
($q=u,d,c,s$), where, for example,
$LR$ refers to an operator with left-handed $\mu$ chirality
and right-handed $q$ chirality.
Lumping all chiralities together, we \cite{kingman} can roughly
summarize the $\Lambda$ values that can be probed at $95\%$ CL. For the
luminosities quoted above:
(a) the $\rts=314\gev$ HERA-analogue machine would be sensitive to
$\Lambda^{\mu u}\sim (0.8-2.0)\tev$ and $\Lambda^{\mu d}\sim (0.7-1.3)\tev$;
(b) the Tevatron machines probe $\Lambda^{\mu u}\sim (7-12)\times \rts$ and 
$\Lambda^{\mu d}\sim (4-8)\times\rts$. In particular, the
$200\gev\otimes 1\tev$ machine probes $\Lambda^{\mu u}$'s $\sim (6-11)\tev$
and $\Lambda^{\mu d}\sim (4-7)\tev$, \ie\ far beyond the HERA machine level.
However, if $\Lambda^{eu}$ and/or $\Lambda^{ed}$ is non-zero 
(as perhaps indicated by the HERA $ep$ excess),
an acceptably small level of FCNC requires
$\Lambda^{\mu u,\mu d}\simeq 0$; only $\Lambda^{\mu c,\mu s}$ 
could be significant in size. Due to the smaller size of the $c$ and $s$
distribution functions in the proton, the $\Lambda^{\mu c}$ ($\Lambda^{\mu s}$)
values that can be probed are typically a factor of 3--4 (1.5--2) smaller than
the $\Lambda^{\mu u}$ ($\Lambda^{\mu d}$) values quoted above.

We turn next to lepto-quarks.
for a first comparison \cite{kingman} of
different colliders we focus on a $+2/3$ charge spin-0 lepto-quark
with $\ell^+ d$ and/or $\ell^+ s$ couplings.
The relevant coupling is defined by
$\call=\lam_{\ell q} L_q\anti q P_\tau \ell$ ($\tau=L$ or $R$).
We take $\br(L_q \to \ell^+ q)=1$ 
and require a 95\% CL for the signal with respect to predicted
background. Table~\ref{leptoq} shows that one can probe the same level of
$\lam_{\ell^+ q}$ ($q=s,d$) at much higher $M_{L_q}$ (or increasingly smaller
$\lam_{\ell^+ q}$ at the same $M_{L_q}$) as the $\mu p$ collider energy
and luminosity increases.  Of course, if there is a lepto-quark with
$\lam_{e^+ d}\neq 0$ (as possibly
hinted by HERA data), then FCNC limits require $\lam_{\mu^+ d}\simeq 0$;
but, $\lam_{\mu^+ s}$ for this same lepto-quark could be non-zero.
Table~\ref{leptoq} shows that at $M_{L_q}=200\gev$, the $\lam_{\mu^+ s}$ value
that can be probed at the $\rts =447\gev$ 
($\rts=894\gev$) Tevatron $\mu p$ collider
is comparable to (much smaller than) the $\lam_{e^+ d}$ value that
can be probed at HERA, despite the fact that the
distribution function for $s$ quarks
in the proton is much smaller than that for $d$ quarks.
Further, it is highly
possible that the second family $\lam_{\mu^+ s}$ coupling
would be larger than the first family $\lam_{e^+ d}$ coupling.

\begin{table}
\normalsize
\def\arraystretch{.75} 
\caption[]{\small
$\lambda_{\ell^+ d}$ ($\lam_{\ell^+ s}$) values required for a 95\% CL signal
with respect to background assuming $\br(L_q\to \ell^+ j)=1$.}
\begin{center}
\begin{tabular}{|c||c|c|c|c|c|c|c|}
\hline
\small
 $M_{L_q}$ (GeV) & 200 & 300 & 400 & 500 & 600 & 700 & 800 \\
\hline
{$\sqrt s$ (GeV)~~~ $L$ (fb$^{-1}$)} &  
\multicolumn{7}{c|} {\quad $\lambda_{\ell^+ d}\times 10^3$ 
($\lambda_{\ell^+ s}\times 10^3$) } \\
\hline
{314 ~~~~~~ 0.1} & 14(73) & $-$ & $-$ & $-$ & $-$ & $-$ & $-$ \\
{447 ~~~~~~~ \phantom{1}2} & 4.5(13) & 10(53) & 55(1130) & $-$ & $-$ & $-$ & $-$ \\
{894 ~~~~~~~ 13} & 2(4) & 3(7) & 4(12) & 6(22) & 9(99) & 16(140) &
45(860) \\
\hline
\end{tabular}
\end{center}
\label{leptoq}
\end{table}

We note that if any
evidence for contact or lepto-quark interactions is found,
it will be very important to look
for the corresponding excess events in
the $\ell^+\ell^-\to q\anti q$ ($\ell=e,\mu$) cross-channels. 

Overall, the discovery reach of the $\mu p$ colliders is quite impressive.
Further, it cannot be stressed too strongly that if evidence for contact
interactions or lepto-quarks is discovered in $ep$ or $\epem$ collisions,
then it will be mandatory to build an analogous muon facility
for $\mu p$ and $\mupmum$ interactions so as to explore the lepton
flavor dependence of the new physics.
\bit
\item{\bf Neutrino Beam Physics}
\eit
The neutrino beams from an energetic $\mu$ beam would be excellent
for $\nu N$ fixed target experiments. One possibility examined
at this workshop was a $250\gev$ $\mu$ beam, which yields $\vev{E_\nu}\sim
178\gev$ and a beam of known composition (\eg\ $\nu_\mu$, $\anti\nu_ e$ for a
$\mu^-$ beam).
For $\nu N$ fixed target experiments, the neutrino flux would be about a
thousand times larger than at present machines.
The large flux implies that good statistics would be obtained using light
targets, allowing a more definitive comparison of
$F_2(x,Q^2)$ in charged and neutral current measurements. Improved measurements
of $xF_3$, $F_2^{\rm charm}$, spin-physics distributions,
and $|V_{ub}|^2$ would all be possible as well.

There are also substantial advantages associated with the neutrino
beams from an energetic muon beam for long-baseline neutrino oscillation
experiments \cite{kayser,geer,mohapatra}. For example, by pointing
a muon storage ring at an appropriate underground detector (Soudan, Gran Sasso,
$\ldots$) it would be possible to
probe $\Delta m^2$ and $\sin^2 2\theta$ neutrino-mixing
parameters that are factors
of 10 and 100-1000, respectively, smaller than can be probed
by the MINOS and MiniBooNE experiments \cite{geer}.
The known composition of the neutrino beam would again be a big asset.
\bit
\item
{\bf R-parity Violating Scenarios}
\eit
If there is $\rslash$ of form
$\lam_{ijk}\what L_L^i\what L_L^j \what{\anti E_R^k} +\lam^\prime_{ijk}
\what L_L^i\what Q_L^j \what{\anti D_R^k}$ (\ie\ baryon number
is conserved and lepton number is violated), then
many new physics signals arise.
(i) $\lampr\neq 0$ allows an
interpretation of the HERA events in which a squark
plays the role of a lepto-quark;
most likely $e^+ d\to \wtil t$ or $\wtil c$.
Even if the HERA excess disappears, the analogous $\mu^+ s\to \wtil c$
(no family transition) and $\mu^+ s\to \wtil t$ couplings could
be much larger, in analogy to standard Yukawa coupling trends.
A $\mu p$ collider would then be a very exciting machine.
(ii) $\lam\neq 0$ would lead to the
possibility of $\epem\to \snu_\tau$ ($\lam_{131}$) and/or $\snu_\mu$
($\lam_{121}$)
and $\mupmum\to \snu_\tau$ ($\lam_{232}$) and/or $\snu_e$ ($\lam_{122}$);
$s$-channel $\epem$ and $\mupmum$ production of a $\snu$
is an exciting prospect.

Sensitivities to the squark couplings are related to those
in the general lepto-quark case (aside from corrections
needed for possibly different branching ratios to the final
state of interest); for example, we would identify
$\lam_{e^+ d}\to \lampr_{1j1}$,
$\lam_{\mu^+ d}\to \lampr_{2j1}$,
$\lam_{e^+ s}\to \lampr_{1j2}$,
$\lam_{\mu^+ s}\to \lampr_{2j2}$,
for $L_q$=$\wtil t$ ($j=3$) or $\wtil c$ ($j=2$). From
the lepto-quark discussion it is apparent that fairly small $\lampr_{2j1,2j2}$
values yield a visible signal for reasonably large squark masses.

Since $\snt$ is probably the lightest of the sneutrinos and since $\lam_{232}$
is probably the largest of the $\lam$'s a muon collider looks especially
interesting for $s$-channel sneutrino production. 
The excellent beam energy resolution of a muon collider would
also be a great advantage. To illustrate \cite{fgh},
assume that only $\lam_{232}$ and $\lampr_{333}$ (surely
the largest of the $\lampr$'s) are non-zero.
For reasonable
superpartner masses, appropriate limits are $\lam_{232} \lsim
0.06$, $\lampr_{333} \lsim 1$, and $\lam_{232} \lampr_{333} \lsim
0.004$. The possible $\snt$ decays are:
(a) $\snt\to \nu_\tau \cnone$ if $\mcnone<\msnt$
(with $\cnone$ in turn decaying via $\rslash$ couplings);
(b) $\snt\to \mu^+\mu^-$ (via $\lam_{232}$); and
(c) $\snt\to b\anti b$ (via $\lampr_{333}$).
$\gamsnt$ tends to be large if $\mcnone<\msnt$, but can be very
small if $\snt\to\nu_\tau\cnone$ is disallowed.
Typical decay widths can be found in Ref.~\cite{fgh}.

What will be the role of the muon collider?
In the most likely case, LHC and/or $\rts=500\gev$
\ellc\ data will reveal the existence of R-parity violation and
yield an approximate determination of $\msnt$.
Expectations for the latter are:
(i) $\dm\sim 100\mev$ if $\mupmum$ and or $b\anti b$
decays are observable; 
(ii) $\dm\lsim 2\gev$ if only $\snt\to\nu_\tau\cnone$ decays
give a substantial number of events.
However, even though we know $\rslash$ is present,
only in special situations will it be possible to
determine the actual magnitude of 
$\lam_{232}$ and $\lam^\prime_{333}$ from this LHC or \ellc\ data.
In contrast, unless $\lam_{232}$ is very small, the muon collider will allow 
an accurate measurement of $\lam_{232}$ and, possibly, also of
$\lam^\prime_{333}$.

The procedure is analogous to that for a Higgs boson.
Once the $\snt$ is observed, we turn to the \muc\
and scan for the precise location. The
cross section depends on $\gamsnt$ and $\sigrts$
as in \Eq{sigmaform} leading to the following scenarios:
(i)
If $\gamsnt$ is as small as is likely if $\msnt<\mcnone$
($\snt\to \nu_\tau\cnone$ forbidden), $\sigsntbar$
is largest if $\sigrts$ is as small as possible. 
Thus, it is best to use $R=0.003\%$ ($L=0.1\fbi/{\rm yr}$) 
and place scan points using intervals of size $2\sigrts$.
(ii)
If $\snt\to\nu_\tau\cnone$ is observed,
$\gamsnt$ will be large and will be dominated by
$\Gamma(\snt\to \nu_\tau\cnone)$ which, in turn, can be computed
from known $\cnone$ properties. Then,
it is most advantageous to use $R=0.1\%$ ($L=1\fbi/{\rm yr}$)
and use a scan interval of ${\rm max}[2\sigrts,\gamsnt]$.

\begin{figure}[h]
\begin{center}
{\epsfig{file=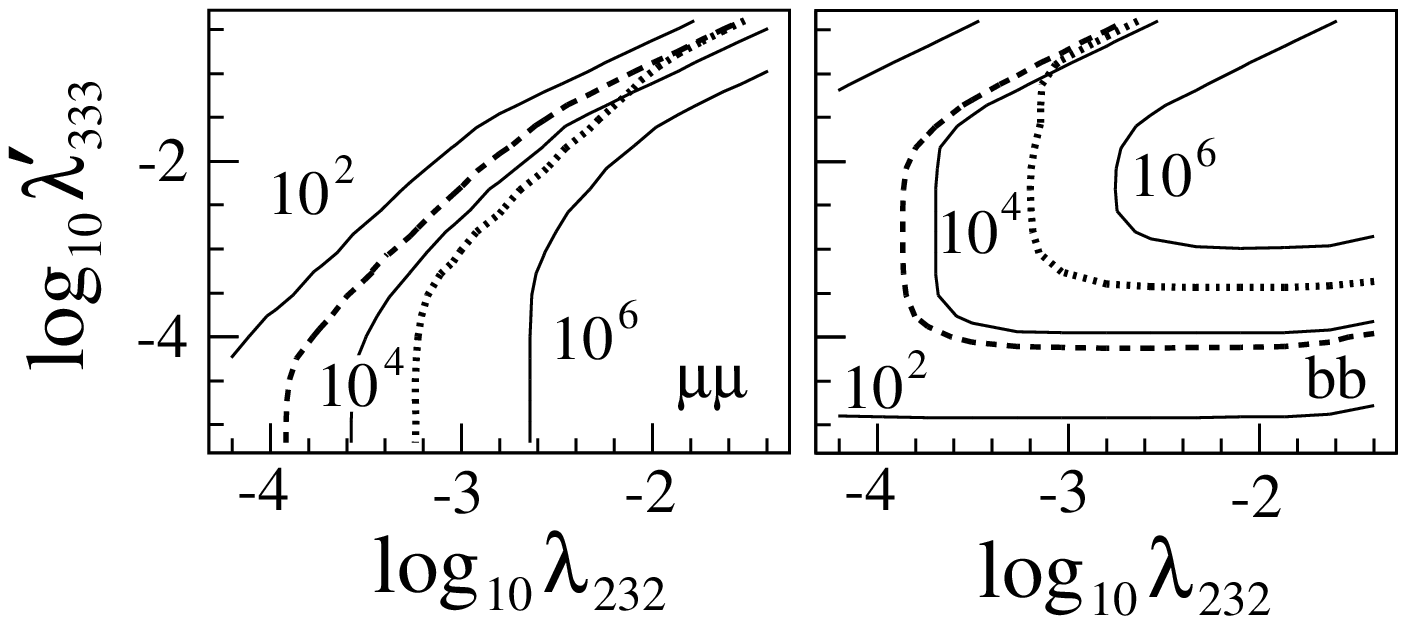,width=3.5in} }
\caption{We consider $\msnt<\mcnone$, $\msnt=100\gev$.
Solid contours are for $\overline\sigma_{\snt}\br(\snt\to X)$ (in fb),
$X=\mu^+\mu^-,b\anti b$. 
The dashed (dotted) contour is the optimistic
(pessimistic scan) $3\sigma$ discovery boundary for
$\ltot = 0.1\fbi$ and $R=0.003\%$.
\label{rates1}}
\end{center}
\end{figure}

\begin{figure}[h]
\begin{center}
{\epsfig{file=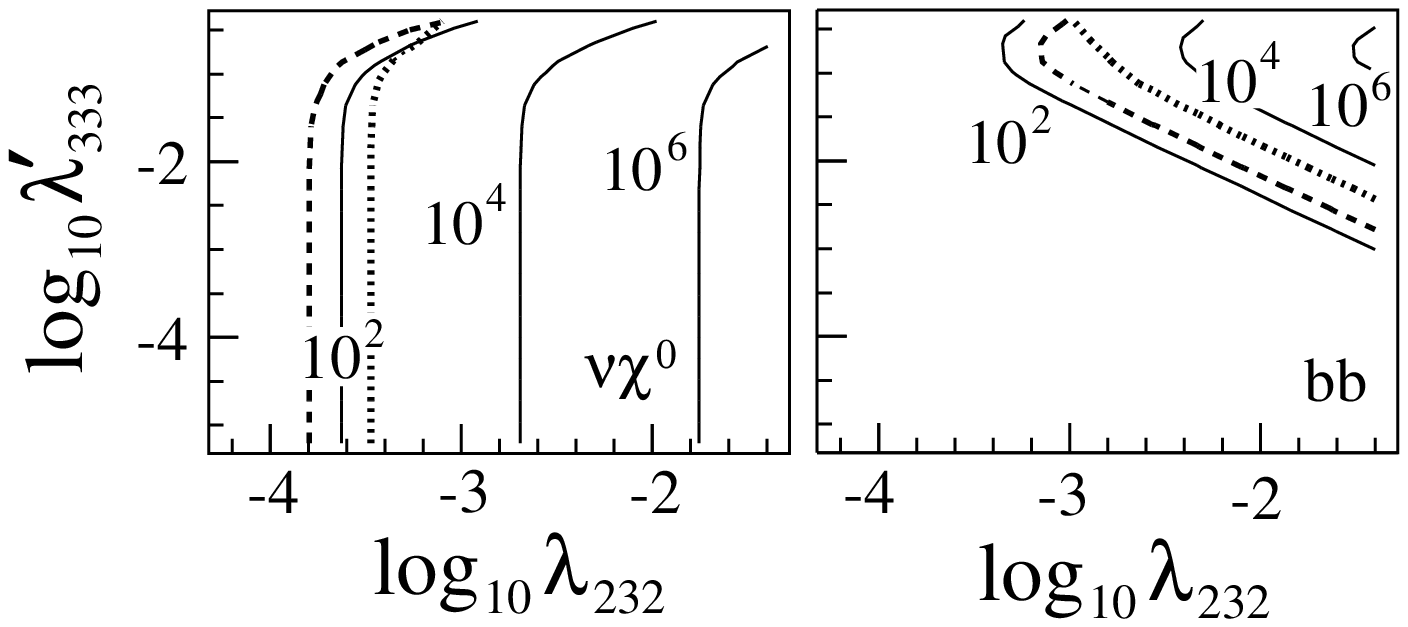,width=3.5in} }
\caption{As in Fig.~\ref{rates1}, but for the $X=\nu\cnone$ 
and $b\anti b$ final states in the $\msnt > \mcnone$ scenario, with
$\msnt = 150\gev$ and a bino-like $\cnone$ with $\mcnone = 100\gev$.
Discovery boundaries assume $\ltot=1\fbi$ at $R=0.1\%$.
\label{rates2}}
\end{center}
\end{figure}
The regions of $(\lam_{232},\lam^\prime_{333})$ parameter 
space for which the $\snt$
could be discovered by the $s$-channel scan over the 
$100\mev$ and $2\gev$ mass windows in cases (i) and (ii)
are illustrated by the dotted contours in Figs.~\ref{rates1} 
and \ref{rates2}, respectively.
The prospects are excellent unless the R-parity violating $\lam_{232}$
coupling is really quite tiny. 

We have estimated
the accuracy with which the R-parity violating couplings can be measured
once the $\snt$ is discovered, assuming 3 years of operation
[$L=0.3\fbi$ in case (i)
and $L=3\fbi$ in case (ii)] distributed at and/or near $\rts=\msnt$.
For $\mcnone>\msnt$ and $\lam_{232}=5\times 10^{-4}$
(which is on the border of the scan discovery regions
and typical of many models),
we find $1\sigma$ errors of $\Delta\lam_{232}/\lam_{232}\sim 2\%-15\%$;
$\Delta\lam^\prime_{333}/\lam^\prime_{333}\sim 10\%-30\%$ is achieved
if $\lam^\prime_{333}$ is not too small.
If $\snt\to\nu_\tau\cnone$, $\lampr_{333}$ must be substantial to be measurable.
However, even if $\lampr_{333}\to 0$, the $\nu_\tau\cnone$ final
state will yield
$\Delta\lambda_{232}/\lambda_{232}\sim 0.9,9,15,25,80\%$ for
$\lambda_{232}=10^{-3},10^{-4},5\times 10^{-5},3\times 10^{-5},
10^{-5}$, respectively.

Finally, we note that the ability to achieve
$R=0.003\%$ is a unique muon collider feature that could allow
one to resolve the $\calo (1\mev)$ splitting between the CP-even and
CP-odd $\snu_\tau$ components that is predicted 
using generic relationships to neutrino
masses and a $\nu_\tau$ mass in the $\gsim 1\mev$ range.
\bit
\item
{\bf\boldmath Probes of Technicolor and Strong $WW$ Scattering}
\eit
The ability of a high energy muon collider
($\rts\gsim 3\tev$) with high $\call$ ($\call\gsim
10^{33}-10^{34}$cm$^{-2}$s$^{-1}$ is anticipated)
to search for heavy technicolor or related resonances
or explore a strongly interacting $WW$ sector has been well documented
\cite{bbghstrong} and will not be reviewed here. Additional work at
this meeting in this area appears in \cite{bhat}. Here we briefly
summarize the ability of a low-energy muon collider to observe
the pseudo-Nambu-Goldstone bosons (PNGB's) of an extended technicolor
theory.  These are narrow states,
that, as noted earlier, need not have appeared at an observable
level in $Z$ decays at LEP. Some of the PNGB's have
substantial $\mupmum$ couplings. Thus, a muon collider search for
them will bear a close resemblance to the light Higgs and R-parity
violating sneutrino cases discussed already.  The main difference
is that, assuming they have not been detected ahead of time, we must search
over the full expected mass range.

The results of PNGB studies at this meeting appear
in Refs.~\cite{dominici} and \cite{lane}.
Here I summarize the results for the lightest $\pzero$ PNGB
as given in Ref.~\cite{dominici}. 
Although the specific $\pzero$ properties employed are those predicted
by the extended BESS model \cite{dominici} , they will be
representative of what would be found in any extended technicolor model for
a strongly interacting electroweak sector. The first point
is that $\mpzero$ is expected to be small; $\mpzero\lsim 80\gev$ 
is preferred in the BESS model. 
Second, the Yukawa couplings and
branching ratios of the $\pzero$ are easily determined.
In the BESS model,
${\cal L}_Y= -i \sum_f\lambda_f \bar f\gamma_5 f \pzero$
with
$\lambda_b=\sqrt{\frac 2 3 }\frac {m_b} v$,
$\lambda_\tau=-\sqrt{6}\frac {m_\tau} v$,
$\lambda_\mu=-\sqrt{6} \frac {m_\mu} v$.
Note the sizeable $\mupmum$ coupling.
The $\pzero$ couplings to $\gamma\gamma$ and $gg$
from the ABJ anomaly are also important.
Overall, these couplings are not unlike
those of a light Higgs boson.  Not surprisingly, therefore,
$\gampzero$ is very tiny: $\gampzero=0.2,4,10\mev$ for $\mpzero=10,80,150\gev$,
respectively, for $N_{TC}=4$ technicolor flavors. For such narrow
widths, it will be best to use $R=0.003\%$ beam energy resolution.

\begin{figure}[h]
\begin{center}
\epsfig{file=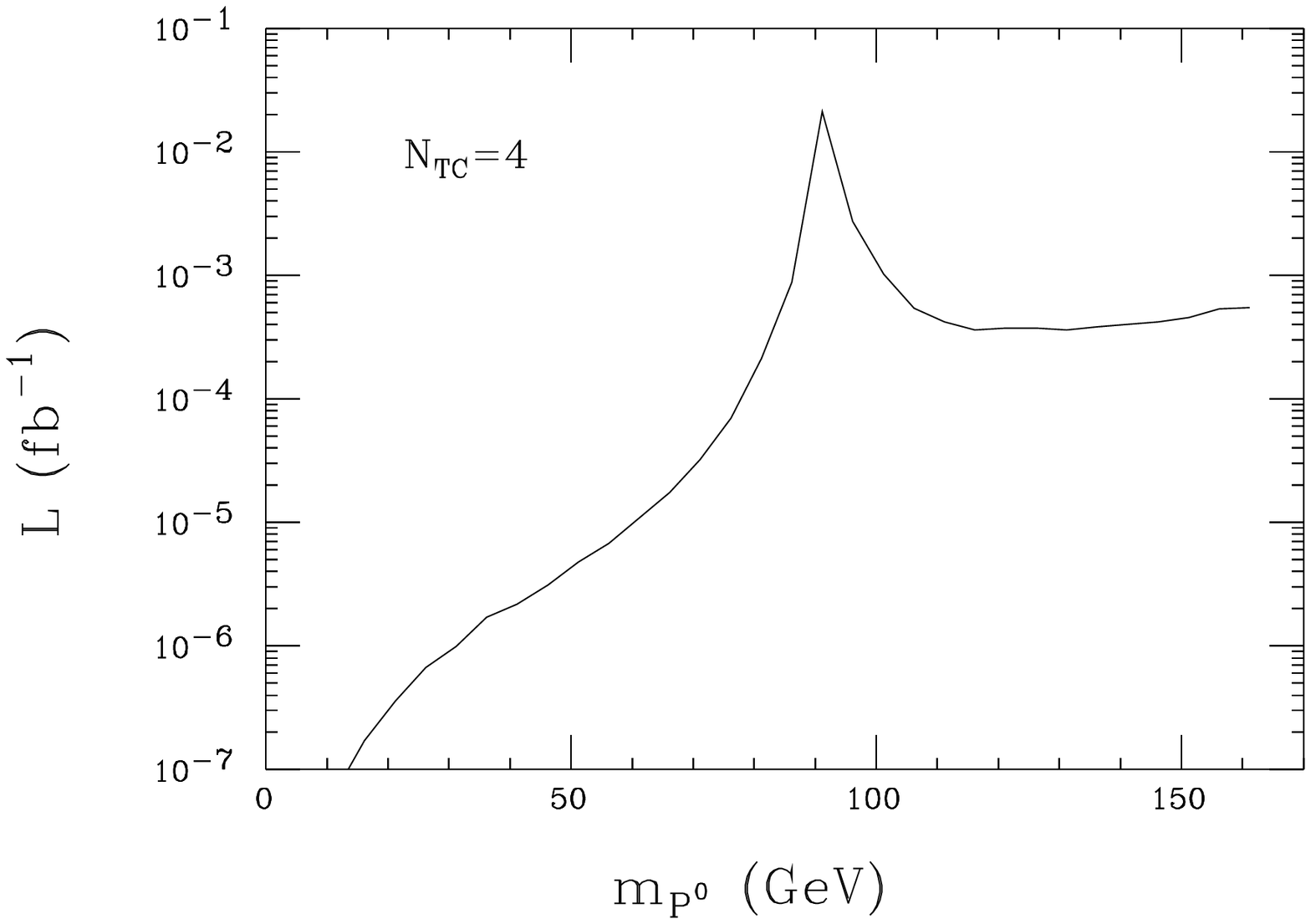,width=3.5in}
\caption{$\ltot$ required 
for a 5$\sigma$ $\pzero$ signal at $\protect\rts=\mpzero$.
\label{figlum}}
\end{center}
\end{figure}

\begin{figure}[h]
\begin{center}
\epsfig{file=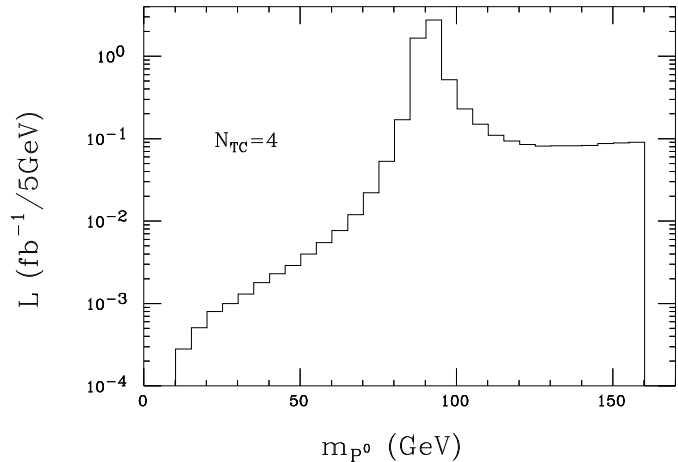,width=3.5in}
\caption{$\ltot$ required to scan indicated 5 GeV intervals
and either discover or eliminate the $\pzero$ at the $3\sigma$
level. \label{lumtable}}
\end{center}
\end{figure}
For the detailed tagging efficiencies \etc\ 
described in \cite{dominici}, the $\ltot$
required to achieve $\sum_k S_k/\sqrt{\sum_k B_k}=5$
at $\rts=\mpzero$, after summing over the optimal selection of
the $k=b\bar b$, $\tau^+\tau^-$, $c\bar c$, and $gg$ channels
(as defined after tagging), is plotted in Fig.~\ref{figlum}. 
Very modest $\ltot$ is needed unless $\mpzero\sim\mz$.
Of course, if we do not have any information regarding the $\pzero$ mass,
we must scan for the resonance. The (very conservative, see \cite{dominici}
for details) estimate for the luminosity required for scanning a 
given 5 GeV interval and either discovering
or eliminating the $\pzero$ in that interval
at the $3\sigma$ level is plotted in Fig.~\ref{lumtable}.
If the $\pzero$ is as light as expected in the extended BESS model, then
the prospects for discovery by scanning would be excellent. For example,
a $\pzero$ lying in the $\sim 10\gev$ to $\sim 75\gev$ mass interval
can be either discovered or eliminated at the $3\sigma$ level with
just $0.11\fbi$ of total luminosity, distributed in proportion to
the luminosities plotted in Fig.~\ref{lumtable}. 
The $\call$ that
could be achieved at these low masses is being studied \cite{palmer}.
A $\pzero$
with $\mpzero\sim\mz$ would be much more difficult to discover
unless its mass was approximately known. A $3\sigma$ scan
of the mass interval from $\sim 105\gev$ to $160\gev$ would require
about $1\fbi$ of integrated luminosity, which is more than
could be comfortably achieved for the conservative $R=0.003\%$ $\call$
values assumed for this workshop.

\section*{Discussion and Conclusions}

There is little doubt that a variety of accelerators will be needed 
to explore all aspects of the physics that lies beyond the Standard Model
and accumulate adequate luminosity for this purpose in a timely fashion.
Certainly, a muon collider (preferably in conjunction with
a $\mu p$ option) would make major contributions to understanding 
any foreseeable type of new physics. It would be of special value in studying 
narrow resonances with $\mupmum$ couplings (such as the SUSY Higgs
bosons) and the lepton flavor dependence of many important classes
of new physics. 
The physics motivations for a muon collider are undeniable and we should
proceed with the R\&D required to assess its viability.
Finding designs that yield the highest
possible luminosity at low energies, while maintaining excellent
beam energy resolution, should be a priority.

\end{document}